\documentclass{pasa}%

\usepackage{graphicx}

\title[CME source regions]{Difference of source regions between fast and slow coronal mass ejections}

\author[B. Filippov]{B. Filippov \thanks{E-mail:
bfilip@izmiran.ru}
\affil{Pushkov Institute of Terrestrial Magnetism,
Ionosphere and Radio Wave Propagation of the Russian Academy of
Sciences (IZMIRAN), \\ Troitsk, Moscow 108840, Russia}%

}%

\jid{PASA}
\doi{10.1017/pas.\the\year.xxx}
\jyear{\the\year}

\usepackage{aas_macros}
\usepackage{hyperref} 
\hypersetup{colorlinks,citecolor=blue,linkcolor=blue,urlcolor=blue}

\hypersetup{draft}

\begin{document}

\begin{frontmatter}
\maketitle

\begin{abstract}
Coronal mass ejections (CMEs) are tightly related to filament eruptions and usually are their continuation in the upper solar corona. It is common practice to divide all observed CMEs into fast and slow ones. Fast CMEs usually follow eruptive events in active regions near big sunspot groups and associated with major solar flares. Slow CMEs are more related to eruptions of quiescent prominences located far from active regions.  We analyze ten eruptive events with particular attention to the events on 2013 September 29 and on 2016 January 26, one of which was associated with a fast CME, while another was followed by a slow CME. We estimated the initial store of free magnetic energy in the two regions and show the resemblance of pre-eruptive situations. The difference of late behaviour of the two eruptive prominences is a consequence of the different structure of magnetic field above the filaments. We estimated this structure on the basis of potential magnetic field calculations.  Analysis of other eight events confirmed that all fast CMEs originate in regions with rapidly changing with height value and direction of coronal magnetic field.  
\end{abstract}

\end{frontmatter}

\section{INTRODUCTION }
\label{sec:intro}

Coronal mass ejections (CMEs) are the most dangerous source of space weather  disturbances. They start suddenly and are in many cases unpredictable. Usually they are recorded after appearance in a field-of-view (FOV) of a space-borne coronagraph or in some cases of a ground-based coronagraph. The relation of CMEs to phenomena observed in the low corona is still under debates. Solar flares were initially considered as drivers of CMEs. However, later it was established that CMEs and flares are separate, while related phenomena \citep{We12}. Whereas most energetic CMEs, as a rule, are associated with big flares, most flares occur independently of CMEs \citep{Ya05,Wa07}. In flare associated CME events, the CME onset typically precedes the associated X-ray flare onset by several minutes \citep{Ha91}.

Statistical study shows that CMEs have the greatest correlation with eruptive prominences (filaments) among other near-surface activity  \citep{Mu79,St91,Ho02,Go03}. Nevertheless, there is a difference in latitudinal distribution of CMEs and eruptive prominences. While at solar maximum both events can happen everywhere around the solar limb, close to minimum CMEs cluster around the solar equator, whereas eruptive prominences originate at latitudes typical for active regions  \citep{Pl02,Lo04}. This discrepancy can be explained by non-radial trajectories of eruptive prominences in the middle corona, which are observed in some well documented events  \citep{Go00,Ho02,Pe12,Pa11,Pa13}  and follow from modeling  \citep{Fi01b,Fi02,Fi16c}. In some cases, eruptive prominences can be traced into the upper corona to become CME bright cores  \citep{Ho81,Il86,Go98}.  Cold material evidently belonging to remnants of eruptive filaments is also detected within interplanetary CMEs (ICMEs), which are interplanetary manifestations
of CMEs \citep{Le10,Wa18}.

The CME speed in the FOV of space-borne coronagraphs  varies in a wide range from tens km s$^{-1}$ to more than 2500 km s$^{-1}$, with an average value of about 500 km s$^{-1}$  \citep{Go04}.  \citet{Sh99}  suggested to separate all CMEs into two types: gradual, or slow CMEs, which usually accelerate within the coronagraph FOV, and impulsive, or fast CMEs, which decelerate during propagation in the corona. CMEs with the speed lower than the average speed can be considered as slow, while the others are fast. Slow CMEs in most cases are associated with filament eruptions. Slow CMEs with persistently weak acceleration were known also as balloon-type events \citep{Sr99,Sr00}. Fast CMEs, in contrast, are usually related to solar flares. However, the separation of CMEs into two groups is rather conventional because parameters of flare-associated and non-flare CMEs considerably overlap  \citep{Vr05} and in most energetic events both flares and filament eruptions are observed  \citep{Sc15}. Many researchers agree that one mechanism is sufficient to explain flare-related and prominence-related CMEs  \citep{Ch03,Fe04}. Numerical simulations \citep{To07}  confirmed the possibility to describe both slow and fast CMEs in a unified manner in a frame of a flux-rope model depending only on the structure of the overlying coronal magnetic field.

In this paper, we analyze ten events initiated by filament eruptions, one part of which produced fast CMEs, while another was followed by slow CMEs. We estimate the initial store of free magnetic energy in all source regions to show the resemblance of pre-eruptive situations. On the basis of potential magnetic field calculations we come to conclusion that the difference of the late behaviour of eruptive prominences is a consequence of the different structure of magnetic fields above the filaments.

\begin{figure*}
		\includegraphics[width=180mm]{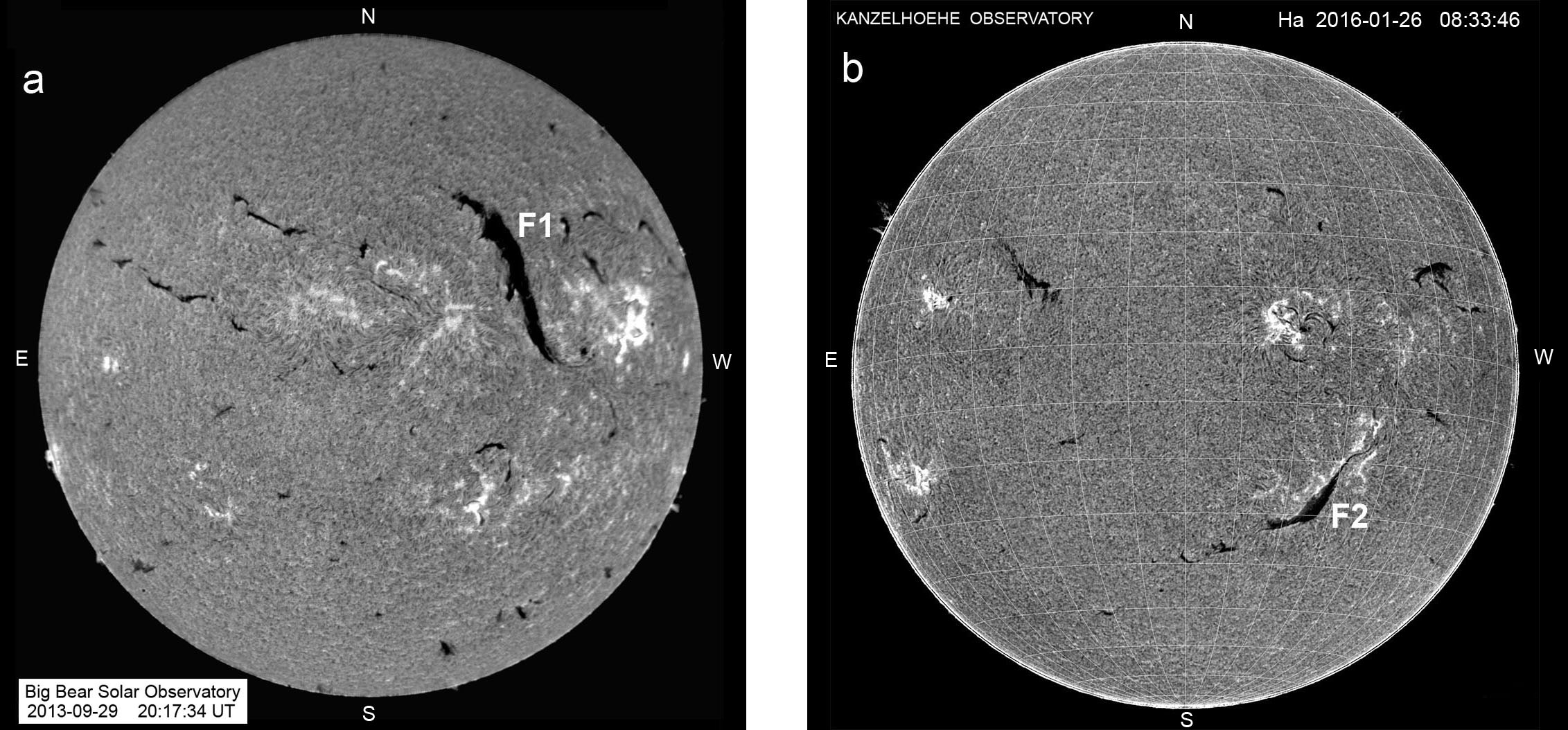}
    \caption{Full disc H$\alpha$ images of the Sun on 2013 September 29 at 20:17:34 UT (a) and on 2016 January 26 at 11:34:15 UT (b). (Courtesy of Big Bear and Kanzelhoehe Solar Observatories). }
    \label{Fig1}
\end{figure*}

\section{Two examples of filament eruptions and CMEs}

\subsection{2013 September 29 event}

  Big quiescent filament erupted after 20:30 UT on 2013 September 29 {\bf (movie1)}. Full disc H$\alpha$ image of the Sun taken at the Big Bear Solar Observatory (BBSO) with the big prominent filament stretched at an angle of about 10$^\circ$ with respect to the south-north direction is shown in Figure \ref{Fig1}(a) just before the start of the eruption. We denote this filament by F1. The ends of the filament deviate to opposite sides from its rather straight body forming the inverse S-shaped structure typical for sigmoidal structures in the northern hemisphere  \citep{Ru96,Pe01}. Filament barbs are right-bearing and thin filament threads deviate clockwise from the filament axis. Both features indicate the dextral chirality of the filament in accordance with the hemispheric chirality rule  \citep{Ma94,Zi97}.

The CME associated with the filament eruption [Figure \ref{Fig2}(a)] appeared above the occulting disc of the Large Angle and Spectrometer Coronagraph (LASCO) C2 \citep{Br95} on board the {\it Solar and Heliospheric Observatory (SOHO)} at 22:12 UT. According to the {\it SOHO}/LASCO CME catalog (http://cdaw.gsfc.nasa.gov/CME\_list/), the CME moved with a linear speed of 1180 km s$^{-1}$ and had a speed of 1165 km~s$^{-1}$ at a distance of 20 $R_\odot$ showing a constant speed. The core of the CME moved within the FOV of LASCO C2 with the averaged speed of 510~km~s$^{-1}$. The mass of the CME is estimated as 2.2$\times$10$^{16}$~g and the kinetic energy as 1.5$\times$10$^{32}$ erg, however these values are marked as rather uncertain. In the catalog, the CME is characterized as a halo CME.

\subsection{2016 January 26 event}

Another filament eruption was observed on 2016 January 26 in the southern hemisphere {\bf (movie2)}. In Figure \ref{Fig1}(b), the filament designated as F2 is shown in the full disc H$\alpha$ filtergram of the Kanzelhoehe Solar Observatory 5 hours before the eruption. The filament is stretched from the south-east to the north-west. Fine structure of the filament reveals the sinistral chirality typical for the southern hemisphere. The eruption starts at about 16:30~UT, and not all length of the filament was involved into the eruption. The eastern and western segments of the filament seem to hold their positions. Only the central section of the filament rises as a big loop. Some filament material falls to the filament ends and to an intermediate footpoint. \citet{Ro18} studied triggers of this filament eruption. They concluded that the filament was destabilized by converging photospheric flows below it, which initiated an ascent of the middle section of the filament up to the critical height of the torus instability.

\begin{figure*}
		\includegraphics[width=180mm]{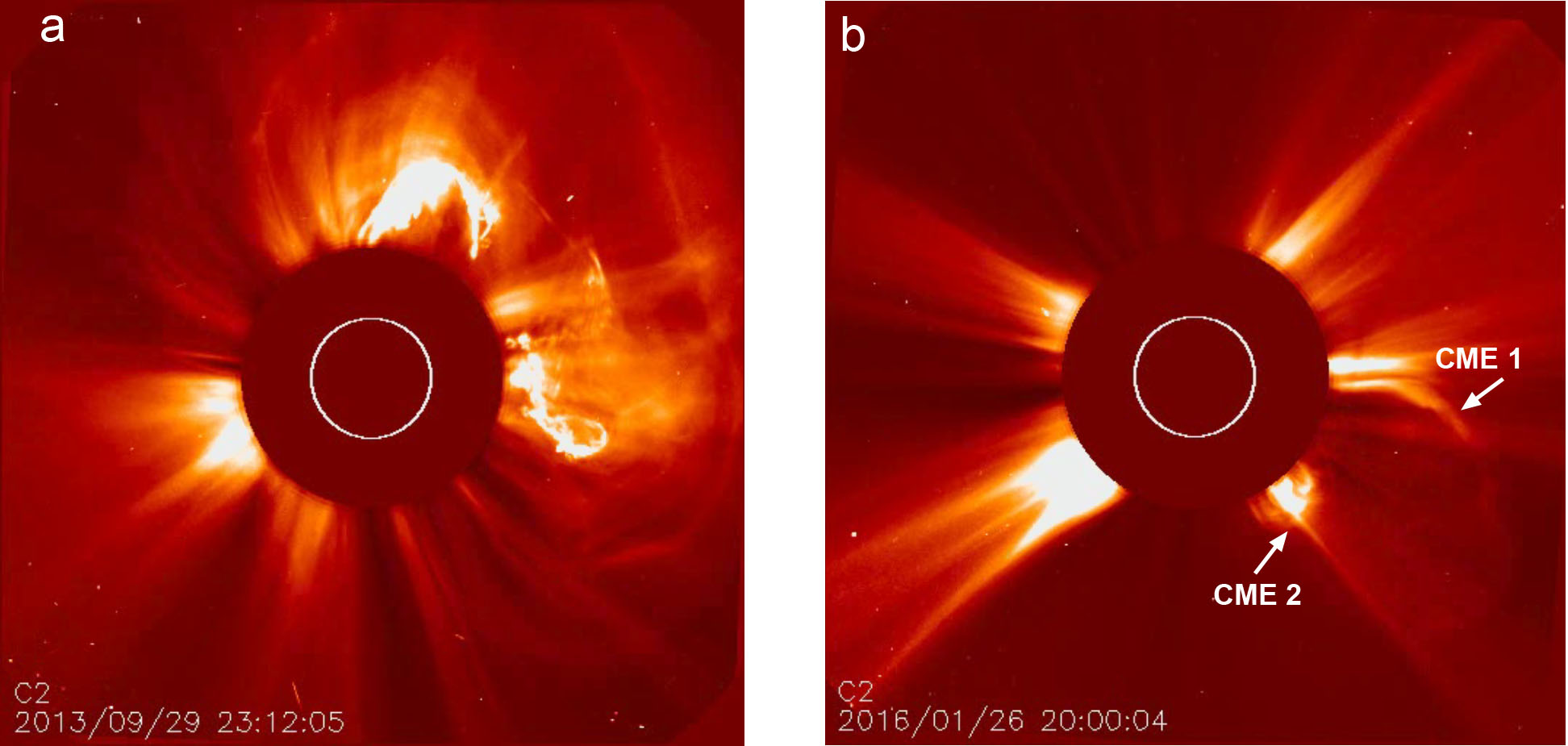}
    \caption{{\it SOHO}/LASCO C2 observations of the CMEs on 2013 September 29 (a) and 2016 January 26 (b). (Courtesy of the {\it SOHO}/LASCO Consortium, ESA and NASA).}
    \label{Fig2}
\end{figure*}

There are two CMEs which may be associated with the filament eruption according their time of appearance and position [Figure \ref{Fig2}(b)]. The first CME appeared at 18:24 UT with the central position angle of 243$^\circ$, a linear speed of 700 km~s$^{-1}$, and a speed of 820 km~s$^{-1}$ at a distance of 20 $R_\odot$ according to the {\it SOHO}/LASCO CME catalog. The mass of the CME is estimated as 1.5$\times$10$^{15}$~g and the kinetic energy as 3.6$\times$10$^{30}$ erg. The second CME appeared at 19:24 UT with the central position angle of 235$^\circ$, a linear speed of 320 km~s$^{-1}$, and a speed of 420 km~s$^{-1}$ at a distance of 20 $R_\odot$ according to the {\it SOHO}/LASCO CME catalog. The core of the CME moved within the FOV of LASCO C2 with the averaged speed of 290 km~s$^{-1}$. The mass of this CME is estimated as 1.1$\times$10$^{15}$~g and the kinetic energy as 5.6$\times$10$^{29}$~erg. 

The observed CMEs could be launched by two independent events in the lower atmosphere, however there was not any other eruptive/flaring phenomenon in the south-west sector of the Earth-side solar hemisphere at convenient time. There was a small eruptive event observed on the far side of the Sun by the {\it  Solar Terrestrial Relations Observatory Ahead (STEREO A)}, which was at that time nearly diametrically opposite the Earth. The eruption was located in the south-west sector of the disc in the framework of  {\it STEREO A}, which was appropriate for the source region of the observed CMEs, but it began at 20 UT, too late to be the source of each CME.

We believe that both CMEs are parts of the same event. The first CME represents the frontal structure, while the second one corresponds to the core of the CME. As usual, the frontal structure moves faster than the core. Linear extrapolation of the height-time plots of both CMEs in the {\it SOHO}/LASCO CME catalog shows the same start time about 18 UT. Of course, this estimation gives a little retarded start time because does not take into account the acceleration of a CME at the beginning of an eruption and the distance from the source region to the limb. Nevertheless, this is in accordance with the beginning of the filament eruption at about 16:30 UT.

\section{Energy of filament electric currents}

\begin{figure*}
		\includegraphics[width=180mm]{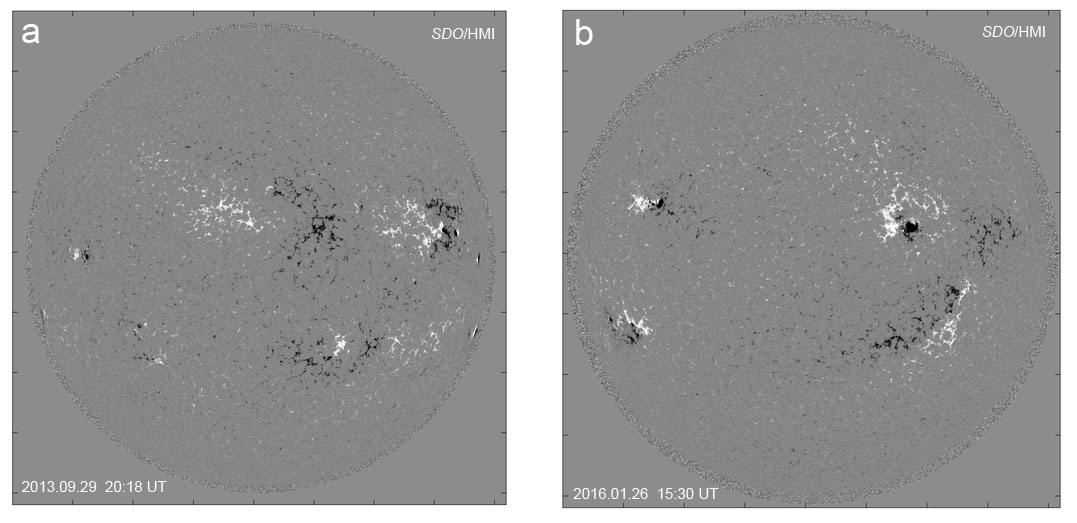}
    \caption{{\it SDO}/HMI images of the line-of-sight magnetic field on 2013 September 29 (a) and on 2016 January 26 (b). (Courtesy of the {\it SDO}/HMI Consortium, ESA and NASA).}
    \label{Fig3}
\end{figure*}

The two filaments were similar in size and erupted in similar ways, but produced very different CMEs. At first we compare the initial conditions of the pre-eruptive filaments. Both filaments were located between large-scale areas of opposite magnetic polarities (compare Figure \ref{Fig1} and Figure \ref{Fig3}). Following a flux rope model of the filament magnetic structure we can estimate the total initial electric currents associated with both filaments and the initial magnetic energies.

In the simplest model with the flux rope considered as a straight linear current, the vertical equilibrium is described as  \citep{va78,Mo87,Pr90}

\begin{equation}
 \frac{I^2}{c^2 h} -  \frac{I}{c} B_t(h) - m g =0,
\end{equation} 

\noindent  where $I$ is the total electric current, $h$ is the height of the electric current above the photosphere, $B_t$ is the horizontal component of the surrounding magnetic field, $m$ is the mass of the tube per unit length, $g$ is the free fall acceleration. Neglecting gravitation when compared with magnetic forces, we can estimate the strength of the total electric current 

\begin{equation}
I =c h B_t.
\end{equation} 

Of course, we cannot obtain both needed values directly from observations but we can estimate them using potential field calculations. We cut some rectangular areas surrounding the filaments from the full-disc magnetograms (Figure \ref{Fig3}) and transform it into array with pixels of equal area. The obtained image looks like if it is observed at the centre of the solar disk. We use the modified array as the boundary conditions for solving the Neumann external boundary-value problem (see  \citep{Fi13} and references therein).

Figure \ref{Fig4} presents the results of potential field calculations. Two top rows show clipped portions of  H$\alpha$ filtergrams and magnetograms  transformed into equal-area-pixel arrays. In the middle row, PILs at different heights are superposed on the H$\alpha$ images. The PILs are drawn taking into account the projection effect, so PILs touching the spines of filaments correspond to their heights \citep{Fi16a,Fi16b}. The height of the filament F1 is between 60 and 84 Mm, while the height of the filament F2 is between 36 and 48 Mm. Figure \ref{Fig4} (d and i) shows the distribution of the decay index  \citep{Ba78,Fi00,Fi01a,Kl06} 

 \begin{equation}
n = - \frac{\partial \ln B_t}{\partial \ln h},
\end{equation} 

\noindent at heights where the contours $n = 1$ touch PILs. These heights can be considered as critical heights $h_c$ from which filaments start to erupt \citep{va78,Fi00,Fi01a,De10,Ol10}.  There are other estimations for the threshold of eruptive instability.  A toroidal current ring becomes unstable if $n = 1.5$ \citep{Ba78,Kl06}. The anchoring of the flux-rope ends in the photosphere \citep{Ol10} and taking into account its finite cross section \citep{De10}  reduce the critical value of the decay index $n_c$ to close vicinity of unity. The majority of studies of the onset of filament eruptions \citep{Fi08,Mc15,Zu14a,Zu14b}
 show that filaments begin to accelerate abruptly when they reach the region with the decay index value close to unity. 

On September 29, the critical height $h_c$ is 82 Mm. It is close to the height estimated from Figure \ref{Fig4}(c). Really, the filament starts to rise soon after the shown moment 20:18 UT. The critical height for the filament F2 is 60 Mm. It is greater than estimated height in Figure \ref{Fig4}(h). However, the filament in this image is shown  at the moment about 5 hours before the eruption and is able to rise slowly to a greater height. Thus, the value of the critical height seems reasonable.      

The bottom row of Figure \ref{Fig4} shows distributions of the horizontal field component $B_t$ at the critical heights. The direction of  $B_t$ shown by arrows corresponds to the direction from positive to negative polarity in the photosphere below the filaments. Maximum values of the horizontal field are about 7 G at the height of 82 Mm on September 29 and 12 G at the height of 60 Mm on January 26. Substituting these values into formula (2) we obtain the strength of the total electric currents as 5.7$\times$10$^{11}$ A and 7.2$\times$10$^{11}$ A, respectively. 

The magnetic energy of an electric current is expressed as \citep{Ta66}

 \begin{equation}
W =  \frac{L I^2}{2 c^2},
\end{equation} 

\noindent where $L$ is the inductance of the circuit. The inductance of the line currents is approximately equal to their lengths. The length of the filament F1 in Figure \ref{Fig4}(a) is about 430 Mm, while the length of the continuous part of the filament F2 in Figure \ref{Fig4}(f) is about 290 Mm. If we take into account the whole length of filaments including thin faint ends and intermediate gaps, F1 is 500 Mm and F2 is 470 Mm long. Then the magnetic energy related to F1 is between 7$\times$10$^{31}$ erg and 8$\times$10$^{31}$ erg. The energy related to F2 is between 7.5$\times$10$^{31}$ erg and 12$\times$10$^{31}$ erg.  Parameters of the filaments F1 and F2 are presented in Tables 1 and 2 along with parameters of other 8 filaments associated with fast and slow CMEs.

Our estimations of the electric current strength and the whole magnetic energy show that both filaments are similar in these parameters characterizing the initial conditions with values for F2 a little greater, but F1 produced a fast CME, while F2 initiated a slow CME. To find the reason for different erupting filament behaviour we should consider the structure of the coronal magnetic field above the filaments where the filaments accelerate.

\begin{table*}
\caption{Eruptive filaments associated with fast CMEs.}
\label{T1}
\begin{tabular}{@{}lcccccccc}
\hline
No & Data & $B_t$, G 
& $h_c$, Mm
& $I$, $10^{11}$ A & $L$, Mm & $W$, $10^{31}$ erg 
& $\alpha$ deg & $v$, km s$^{-1}$
\\
\hline
1 & 2011/03/19 & 4 & 45 & 1.8
& 150 & 0.25 & 230 & 1100 \\
2 & 2011/10/27 & 1.5 & 54 & 0.8
& 450 & 0.14 & 70 & 570 \\
3 & 2012/06/23 & 10 & 48 & 4.8
& 275 & 3.2 & 170 & 1260 \\
4 & 2012/08/23 & 3 & 36 & 1.1
& 360 & 0.22 & 150 & 600 \\
5 & 2013/09/29 & 7 & 82 & 5.7
& 430 & 7 & 150 & 1180 \\
\hline
\end{tabular}
\end{table*}

\begin{table*}
\caption{Eruptive filaments associated with slow CMEs.}
\label{T2}
\begin{tabular}{@{}lcccccccc}
\hline
No & Data & $B_t$, G 
& $h_c$, Mm
& $I$, $10^{11}$ A & $L$, Mm & $W$, $10^{31}$ erg 
& $\alpha$ deg & $v$, km s$^{-1}$
\\
\hline
6 & 2013/06/23 & 4 & 78 & 3.1
& 330 & 1.6 & 80 & 260 \\
7 & 2013/08/14 & 3 & 90 & 2.7
& 240 & 0.9 & 8 & 320 \\
8 & 2013/08/16 & 1.5 & 120 & 1.8
& 660 & 1.1 & 12 & 370 \\
9 & 2013/09/23 & 5 & 48 & 2.4
& 250 & 0.7 & 3 & 290 \\
10 & 2016/01/26 & 12 & 60 & 7.2
& 290 & 7.5 & 20 & 320 \\
\hline
\end{tabular}
\end{table*}

\section{Structure of the potential magnetic field in the source regions}

\begin{figure*}
		\includegraphics[width=135mm]{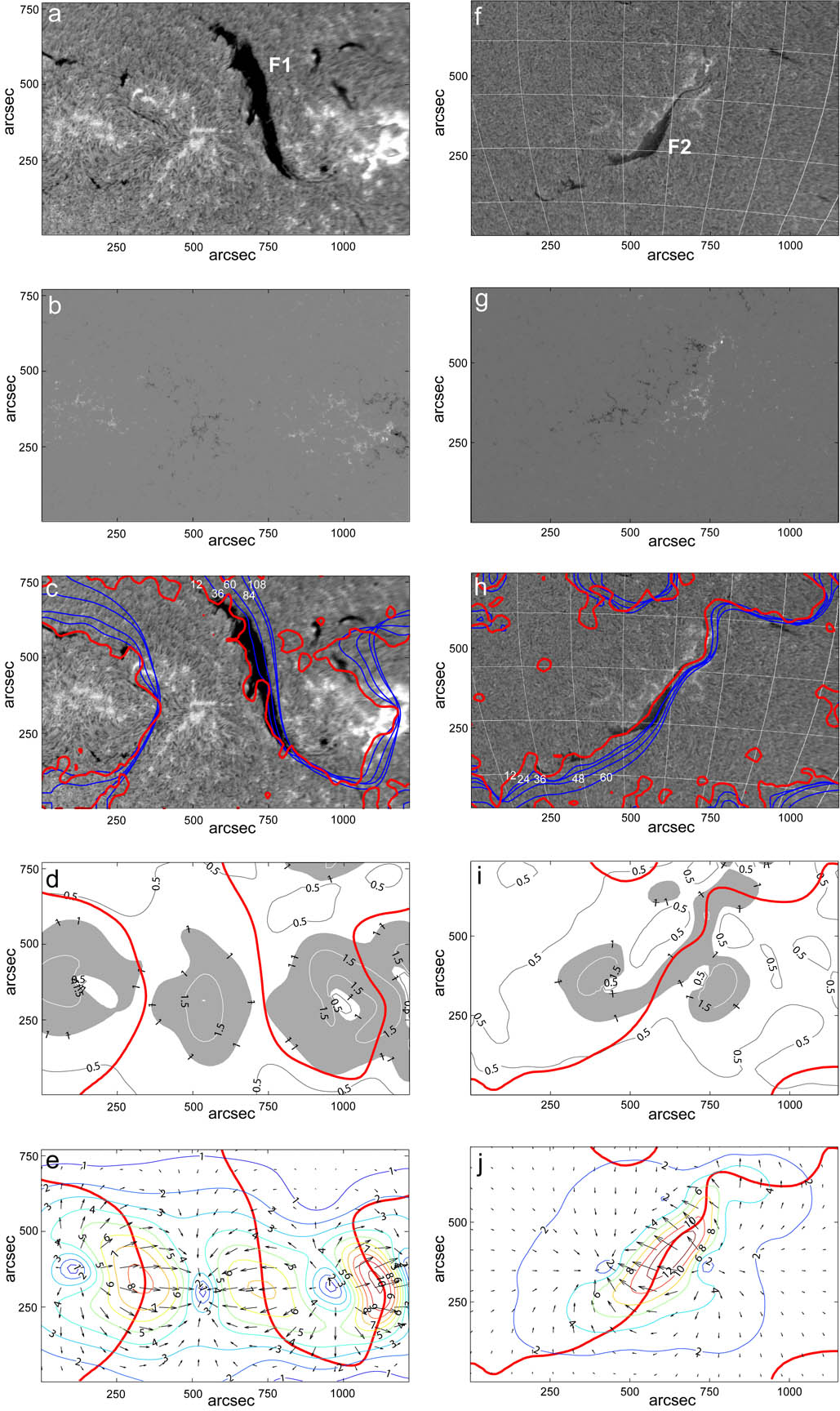}
    \caption{Left column: the fragment of the H$\alpha$ filtergram on 2013 September 29 at 20:18 UT (a), the corresponding fragment of the magnetogram (b), PILs at different heights superposed on the H$\alpha$  filtergram (c), the distribution of the decay index at the height of 82 Mm (d), the distribution of the horizontal field component at the height of 82 Mm (e). Right column: the same for 2016 January 26 at 11:34 UT and the height of 60 Mm. }
    \label{Fig4}
\end{figure*}

\begin{figure*}
		\includegraphics[width=135mm]{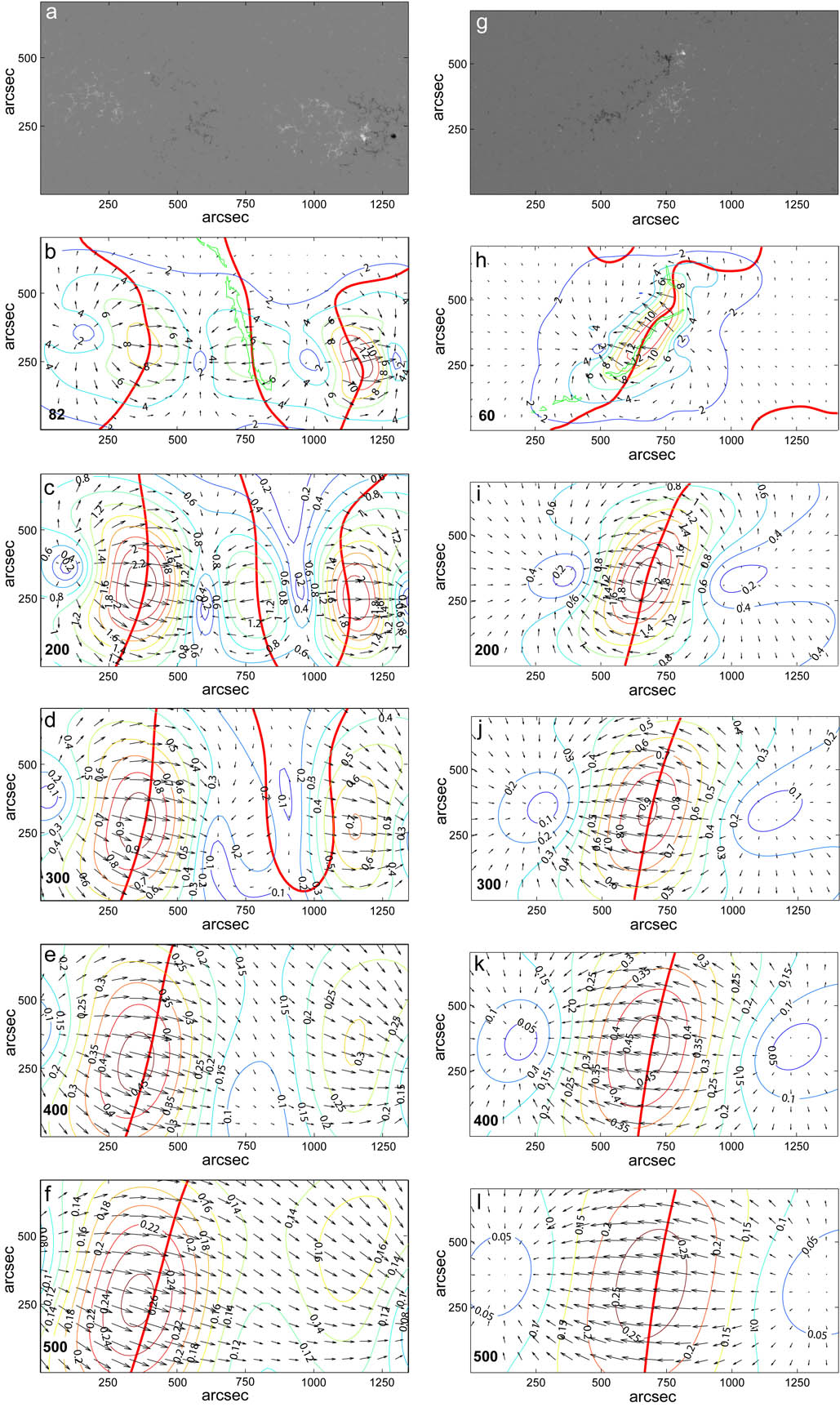}
    \caption{Left column: the fragment of the {\it SDO}/HMI  magnetogram on 2013 September 27 at 20:12 UT (a) and the distribution of the horizontal field component at different heights (b) - (f). Right column: the same for 2016 January 24 at 08:13 UT. Thick red lines show PILs at the corresponding heights. Green contours in the frames (b) and (h) indicate positions of filaments. }
    \label{Fig5}
\end{figure*}

To analyze the structure of the coronal magnetic field above the filaments we use magnetograms taken two days before the eruptive events when the regions were close to the central meridian.  The large-scale structure of the field does not change significantly from day to day, while measurements of the magnetic field in the region when it is near the central meridian are more accurate and reliable. The results are shown in Figure \ref{Fig5}. The top row shows clipped portions of magnetograms used as boundary conditions. The second row presents distributions of the horizontal field component $B_t$ at the critical heights which are practically identical to Figure \ref{Fig4}(e, j) for the days of eruptions. The position of the filaments is shown as green contours. At the heights of 200 and 300 Mm, the horizontal field above the filament F2 is 2-4 times greater than that above the filament F1. At the heights of 400 and 500 Mm, the field above the filament F2 is still slowly decreasing, while the field above the filament F1 changes direction to opposite. The horizontal field above 350 Mm does not hold the filament more but pushes it away from the solar surface. The filament receives additional acceleration and transforms into the fast CME. Despite the smaller initial electric current, the filament F1 is surrounded by a weaker holding magnetic field which decreases rapidly with height and changing the direction to opposite becomes an accelerating field.

The difference in the coronal magnetic field behaviour with height is based on different distributions of photospheric fields. The region near the filament F2 has two large-scale areas of opposite polarities [Figure \ref{Fig5}(g)]. Thus, the structure of the coronal field can be considered as more or less dipolar. The field falls with height not faster than inverse cubic distance from the centre of an effective dipole. The photospheric field around the filament F1 is more complicated. It contains at least four magnetic cells [Figure \ref{Fig5}(a)] less spacious than cells in Figure \ref{Fig5}(g). The coronal magnetic field is more like quadrupolar one and therefore decreases with height faster than dipolar field. It can contain a null point at some height and can change direction to opposite on the other side of the null as we see in Figure \ref{Fig5}.

\begin{figure*}
		\includegraphics[width=180mm]{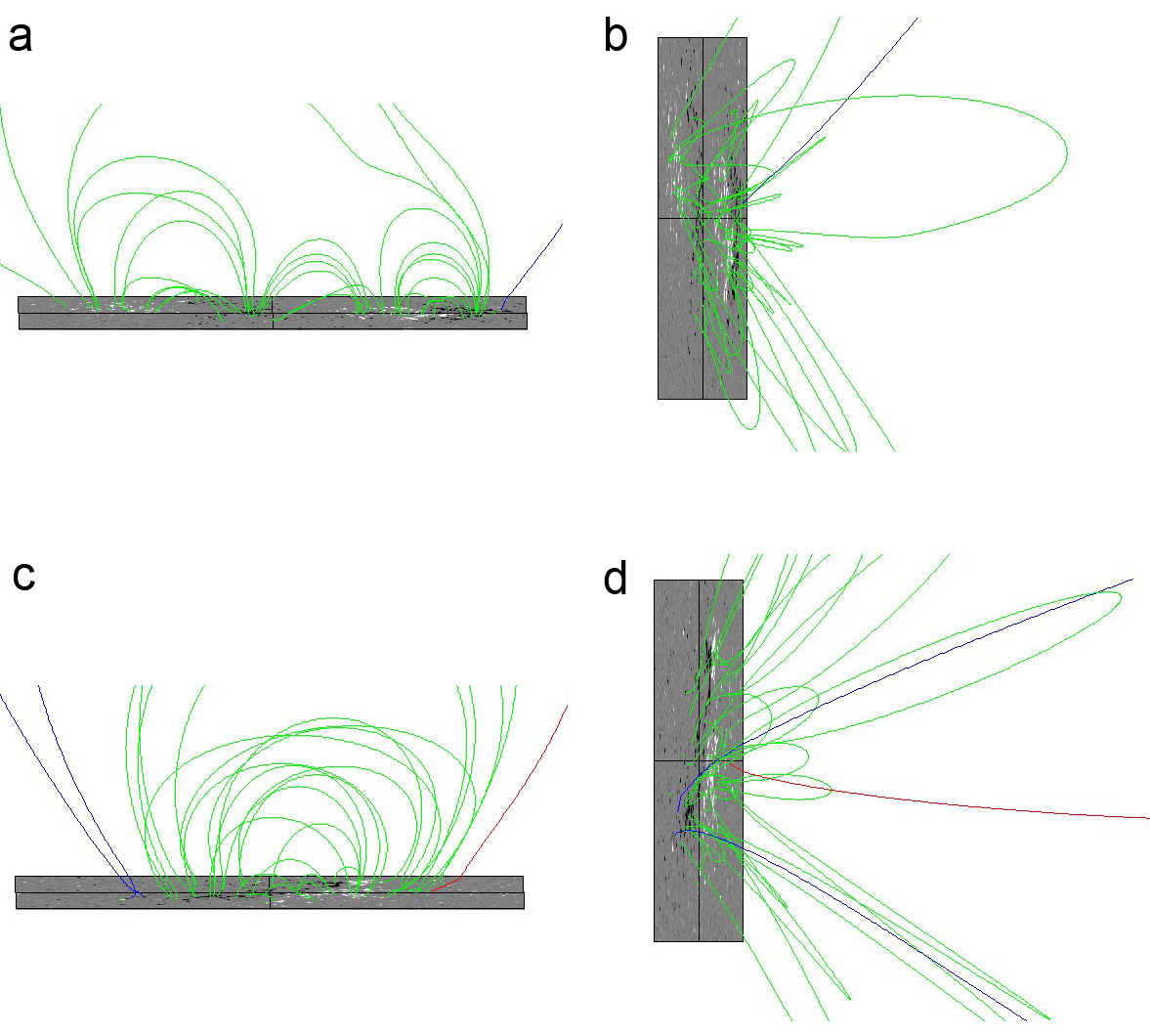}
    \caption{Structure of potential field lines above the filaments F1 (upper row) and F2 (bottom row). }
    \label{Fig6}
\end{figure*}

 Figure \ref{Fig6} shows the structure of potential field lines above the filaments F1 and F2. Left panels present the view form the south,  right panels show the view from the east, as it is expected at the western limb. Field lines above the filament F2 are similar to a simple bipolar arcade, while in the southern view of the region with the filament F1 the quadrupolar structure is clearly visible. A saddle-like geometry above the central loop system indicates the presence of a null point. 

We analyzed additionally four filament eruptions associated with fast CMEs and four eruptions associated with slow CMEs. The results are presented in Tables 1 and 2. The third column presents values of the horizontal component of potential field $B_t$ at the critical height $h_c$ shown in the next column.  The sixth column presents filament lengths measured in filtergrams. Estimations of the electric current strength $I$ and magnetic energy $W$ according expressions (2) and (4) are shown in the fifth and seventh columns. The eighth column presents angles $\alpha$ between the horizontal directions of the potential magnetic field at the heights of 10 Mm and of 600 Mm. In the last column,  linear speeds of the associated CMEs from the {\it SOHO}/LASCO CME catalog are presented. The parameters of the filaments F1 and F2 are shown in the last lines of both tables. On account of many uncertainties in data, all values in tables should be considered as estimations with errors no less than 50\%. 

\begin{figure*}
		\includegraphics[width=180mm]{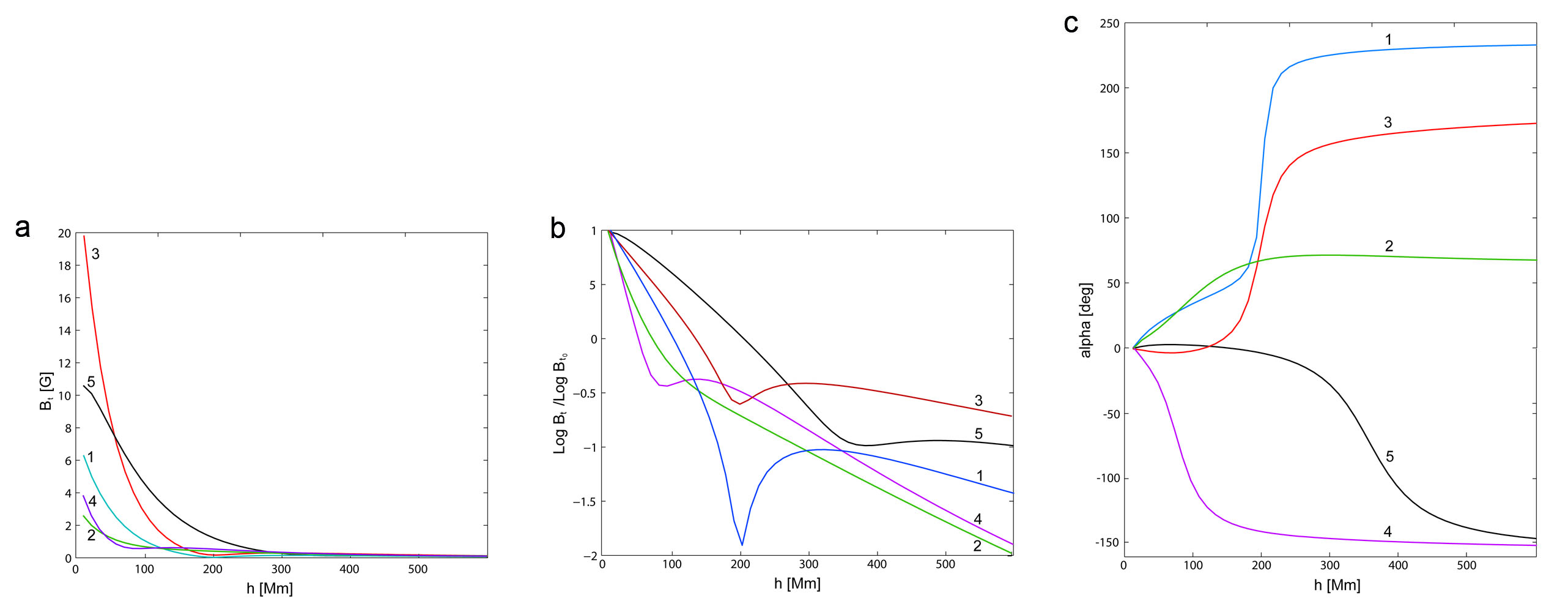}
    \caption{Height profiles of the horizontal component $B_t$ of the potential magnetic field (a), the value of ${\text{Log}} B_t/{\text{Log}} B_{t_0}$, and the angle of the horizontal field rotation $\alpha$ above filaments, which initiated fast CMEs. }
    \label{Fig7}
\end{figure*}

\begin{figure*}
		\includegraphics[width=180mm]{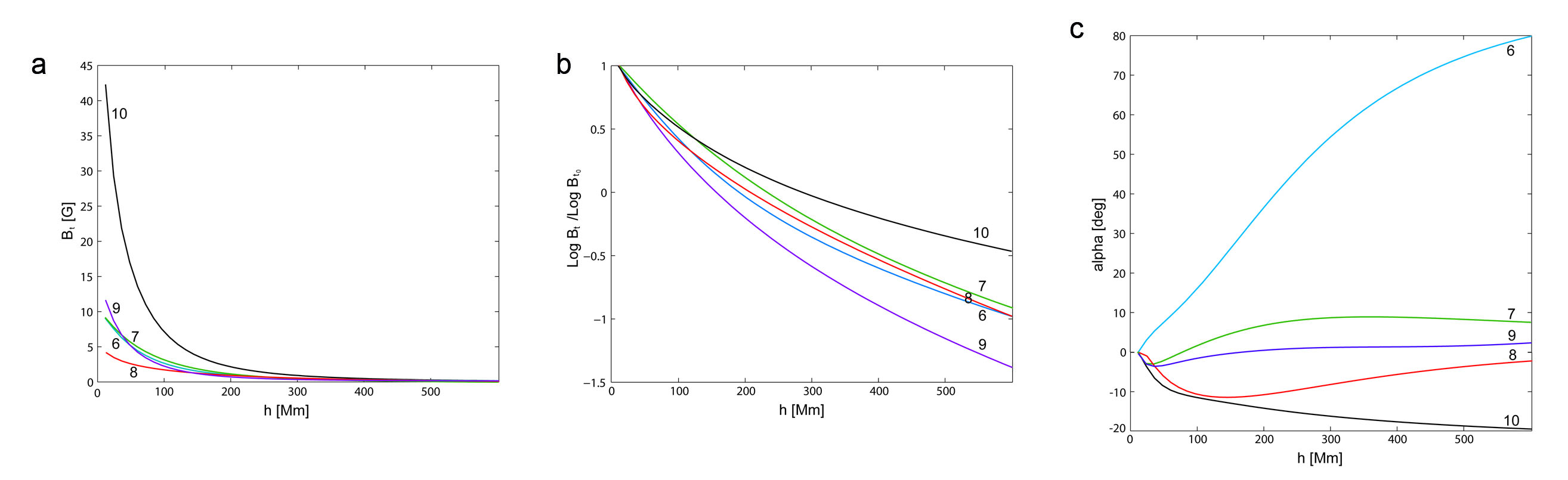}
    \caption{The same as in Figure  \ref{Fig7} foe regions producing slow CMEs.}
    \label{Fig8}
\end{figure*}

The behaviour of the potential magnetic field above the filaments are shown in Figures \ref{Fig7} and  \ref{Fig8} for fast and slow CMEs, respectively. The profiles are calculated along the radial direction starting from the height of 10 Mm near the centre of a filament up to the height of 600 Mm. Every curve is labeled with a figure corresponding to the number of event in Tables 1 and 2. The panels \ref{Fig7}(a) and  \ref{Fig8}(a) show the decrease of horizontal field with height. The behavior of the field is better seen in the panels \ref{Fig7}(b) and  \ref{Fig8}(b) presenting the value ${\text{Log}} B_t/{\text{Log}} B_{t_0}$, where $B_{t_0}$ is the value at theheight of 10 Mm. The panels \ref{Fig7}(c) and  \ref{Fig8}(c) show the rotation of the horizontal field with height. 

Comparison of Table 1 and Table 2 does not demonstrate significant difference in values of electric current and magnetic energy for filaments associated with fast and slow CMEs, while they vary from event to event in each family. However, the values of $\alpha$ in the eighth column are quite different in the two tables. It is clearly visible in Figures \ref{Fig7}(c) and  \ref{Fig8}(c). Rotation of the field to an angle of order of 180$^\circ$ indicates the presence of a quadrupolar structure with a null point. The influence of the nearby null point is manifested in dips on the curves in Figure \ref{Fig7}(b) suggesting the reduction of the field in the vicinity of the null. Only one curve has no a dip. This curve corresponds to the not very fast event (570  km s$^{-1}$), nevertheless the field slopes down faster than any one in  Figure \ref{Fig8}(b). 

The results of our analysis are consistent with the numerical simulations \citep{To07}  of flux-rope eruptions in bipolar and quadrupolar active regions. \citet{To07}  showed that the accelerations profile of the erupting flux rope depends on the steepness of the coronal field decrease with height. The fastest CMEs are expected in most complex active regions. 

There are several simplifications used in our analysis. They, of course, limit the accuracy of some quantitative results. We do not take into account the internal structure of flux ropes containing the filaments and use a simple model with a straight linear current. It seems reasonable because we analyze the equilibrium of the flux rope as a whole and correlate the axis of the flux rope with the filament spine. We consider the photospheric boundary as a flat surface, which is not the case for such large areas. Thus the contribution of sources from the periphery of the areas is accounted not very correctly. Nevertheless, our calculations are more or less correct for the central part of the area up to heights less than the width of the box. 

\section{Summary and conclusions}

 We analyzed ten filament eruptions, one part of which was associated with fast CMEs, while the other was followed by slow CMEs. Particular attention has been given to two  big long-living quiescent filaments nearly of the same size located far from active regions. First eruption happened on 2013 September 29 in the northern hemisphere. It started at 20:30 UT as slow rising of the filament (F1) with continues and increasing acceleration. The eruption produced a big halo CME with the frontal structure moving in the FOV of {\it SOHO}/LASCO with a constant speed of about 1200 km s$^{-1}$. The core of the CME moved within the FOV of LASCO C2 with the averaged speed of 510 km s$^{-1}$. The other filament (F2) erupted on 2016 January 26 at about 16:30 UT in the southern hemisphere. It was associated with a slower CME. The frontal structure of the CME moved with a speed changing from 700 to 800 km s$^{-1}$. The core of the CME propagated within the FOV of LASCO C2 with the averaged speed of 290 km s$^{-1}$. As usual the frontal structure moves faster than the core because the flux rope, which forms a CME, moves translationally and in addition simultaneously expands. 

Our estimations of the electric current strength and the whole magnetic energy show that both filaments are similar in these parameters characterizing the initial conditions with values for F2 a little greater. However,  F1 produced a fast CME, while F2 initiated a slow CME. We ascribe the difference of the  late behaviour of the two eruptive prominences to the different structure of magnetic field above the filaments.  The structure of the coronal field above the filament F2 can be considered as more or less dipolar. The field falls with height not faster than inverse cubic distance from the centre of an effective dipole. The field retards the filament acceleration even at great heights. The coronal magnetic field above the filament F1 is more like quadrupolar one and therefore decreases with height much faster than the dipolar field. It contains a null point at some height and changes direction to opposite on the other side of the null to create additional accelerating force for the filament ascending. 

Similar structures were found in other four regions, which produced  fast CMEs. Most conspicuous feature is the changing  of the horizontal field direction to nearly opposite with the increase of height. Such behavior is natural, if one moves near a magnetic null point. The results of our analysis are consistent with the numerical simulations \citep{To07}  of flux-rope eruptions in bipolar and quadrupolar active regions.

\begin{acknowledgements}
The author thanks the Big Bear and Kanzelhoehe Solar Observatories, {\it SOHO}/LASCO, {\it SDO}/AIA, {\it SDO}/HMI science teams for the high-quality data supplied. The author is grateful to the anonymous referee whose suggestions and comments significantly improved this paper.
\end{acknowledgements}

\bibliographystyle{pasa-mnras}
\bibliography{Refer}

\end{document}